\documentclass[,final]{aipproc}
\newcommand \beq{\begin{eqnarray}}
\newcommand \eeq{\end{eqnarray}}
\newcommand \bea{\begin{eqnarray}}
\newcommand \eea{\end{eqnarray}}
\def\simge{\mathrel{
       \rlap{\raise 0.511ex \hbox{$>$}}{\lower 0.511ex \hbox{$\sim$}}}}
\def\simle{\mathrel{
       \rlap{\raise 0.511ex \hbox{$<$}}{\lower 0.511ex \hbox{$\sim$}}}}

\layoutstyle{6x9}

\begin{document}

\title{Neutron stars and quark matter}

\classification{\texttt{97.60.Jd, 26.60.+c, 12.38.Mh}}


\author{Gordon Baym}{
  address={Department of Physics, University of Illinois at Urbana-Champaign \\
1110 W. Green Street, Urbana, IL, 61801 USA}
}

\begin{abstract}

    Recent observations of neutron star masses close to the maximum predicted
by nucleonic equations of state begin to challenge our understanding of dense
matter in neutron stars, and constrain the possible presence of quark matter
in their deep interiors.

\end{abstract}

\maketitle

\section{Introduction}

\begin{figure}[ht]
  \caption{ Schematic cross section of a neutron star.}
\includegraphics[height=.4\textheight]{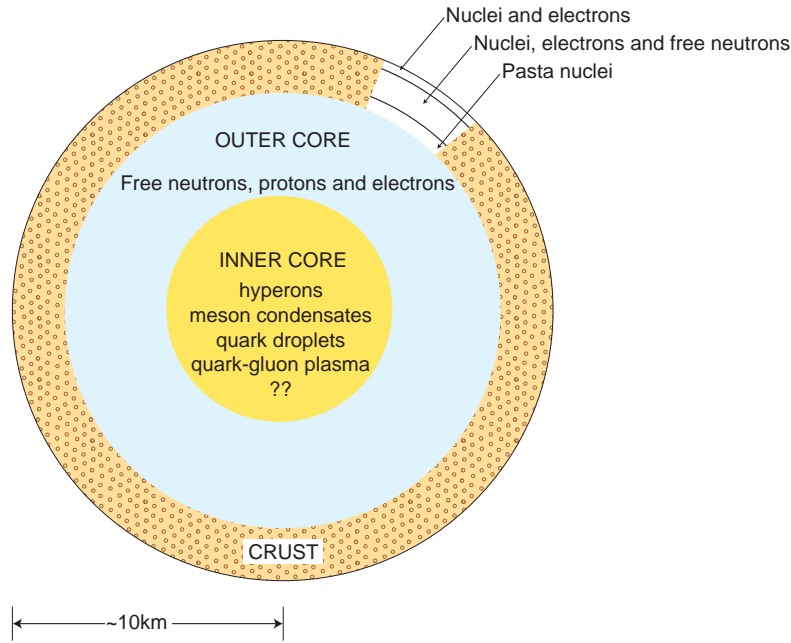}
\label{crosssecn}
\end{figure}

    Neutron stars -- highly compact stellar objects with masses $\sim$ 1-2
$M_{\odot}$ (solar masses), radii of order 10-12 km, and temperatures well
below one MeV -- are natural laboratories to study cold ultradense matter
\cite{review}.  Indeed, the inner cores of neutron stars are the only known
sites where one could expect degenerate quark matter in nature.  Figure
\ref{crosssecn} shows the cross section of a neutron star interior.  The mass
density, $\rho$, increases with increasing depth in the star.  The crust is
typically $\sim$1 km thick, and consists, except in the molten outer tens of
meters, of a lattice of bare nuclei immersed in a sea of degenerate electrons,
as in a normal metal.  The matter becomes more neutron rich with increasing
density, a result of the increasing electron Fermi energy favoring electron
capture on protons, $e^- + p \to n + \nu_e$.  Beyond the {\it neutron drip}
point, $\rho_{drip}\sim 10^{11} $g/cm$^3 (=2\times10^{-4}$ fm$^{-3}$), the
matter becomes so neutron rich that the continuum neutron states begin to be
filled, and the still solid matter becomes permeated by a sea of free neutrons
in addition to the electron sea.  At a density of order half nuclear matter
density, $n_0 \simeq 0.16 $fm$^{-3}$, the matter dissolves into a uniform
liquid composed primarily of neutrons, plus $\sim$5\% protons and electrons,
and a sprinkle of muons.

    The nature of the extremely dense matter in the cores of neutron stars,
while determining the gross structure of neutron stars, e.g., density profiles
$\rho(r)$, radii $R$, moments of inertia, and the maximum neutron star mass,
$M_{max}$, remains uncertain.  Scenarios, from nuclear and hadronic matter, to
exotic states involving pionic \cite{pion} or kaonic \cite{kaon} Bose-Einstein
condensation, to bulk quark matter and quark matter in droplets, including
superconducting states, as well as strange quark matter, have been proposed.
Ultrarelativistic heavy ion collision experiments at RHIC, and soon at ALICE
and CMS at the LHC, probe hot dense matter, from which one can gain hints of
the properties of cold matter.  The uncertainies in the properties of matter
at densities much greater than $n_0$ are reflected in uncertainties in
M$_{max}$, important in distinguishing possible black holes from a neutron
stars by measurement of their masses, and in inferring whether an independent
family of denser quark stars, composed essentially of quark matter, can exist.

\section{Nuclear matter in the interior}

    The properties of the liquid near $n_0$ can be readily determined by
extrapolation from laboratory nuclear physics.  The most reliable equations of
state of nuclear matter in neutron stars are based on extracting
nucleon-nucleon interactions from pp and pn scattering experiments at energies
below $\sim$ 300 MeV, constrained by fitting the properties of the deuteron,
and solving the many-body Schr\"odinger equation numerically via variational
techniques to find the energy density as a function of baryon number, e.g.,
\cite{WFF,akmal}.  The most complete two-body potential is the Argonne A18
(with 18 different components, such as central, spin-orbit, etc., of the
interactions).

    Two-body potentials predict a reasonable binding energy of nuclear matter;
however the calculated equilibrium density is too high.  Similarly, two-body
potentials fail to produce sufficient binding of light nuclei \cite{pieper}.
The binding problems indicate that one must take into account intrinsic
three-body forces acting between nucleons, such as the process in which two of
the nucleons scatter becoming internally excited to an intermediate isobar
state ($\Delta$) while the third nucleon scatters from one of the isobars.
The three-body forces must increase the binding in the neighborhood of
$n_0$, but, to avoid overbinding nuclear matter, they must become repulsive
at higher densities.  This repulsion leads to a stiffening of the equation of
state of neutron star matter at higher densities over that computed from
two-body forces alone.

\begin{figure}[ht]
\includegraphics[height=.32\textheight]{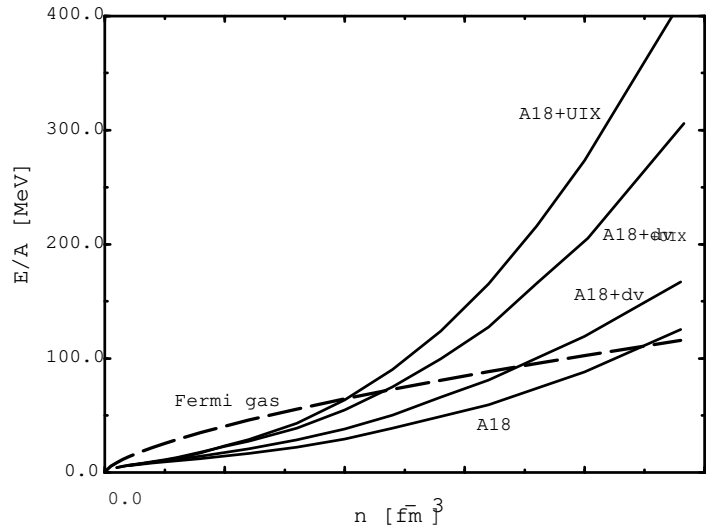}
\caption{Energy per baryon of pure neutron matter as a function of baryon
density, $n$, calculated with the A18 two-body potential with and without the
Urbana IX (UIX) three-body potential, and lowest order relativistic
corrections, $\delta v$.  From \cite{akmal}.
\label{akmal-pnm}}
\end{figure}

\begin{figure}[ht]
\includegraphics[height=.25\textheight]{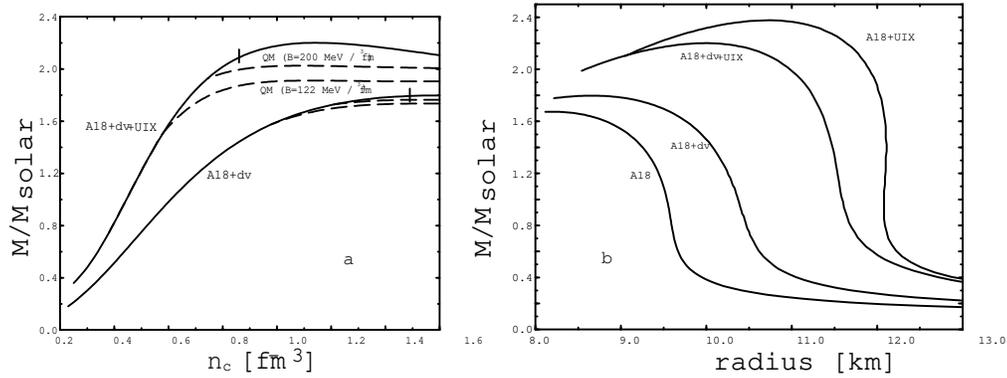}
\caption{a)  Neutron star mass vs. central baryon density for the equations of
state shown in Fig.~\ref{akmal-pnm}, including beta equilibrium.  The curves
labelled QM show the effect of allowing for a transition to quark matter
described in the simple MIT bag model, with bag constants $B$ = 122 and 200
MeV/fm$^3$.  b) Mass vs. radius of neutron stars for the same models.
\label{akmal-structure}}
\end{figure}

    Figure~\ref{akmal-pnm} shows the energy per baryon of neutron matter as a
function of baryon density \cite{akmal} with the A18 two-body potential, and
Urbana UIX three-body potential, together with relativistic boost corrections
($\delta v$), accurate to order $(v/c)^2$.  This equation of state, taking
into account all two-nucleon data, and data from light nuclei, is currently
the best available for $n\simge n_0$.  (Nuclear equations of state based on
the more accurate ``Illinois" three-body potentials \cite{IL3b} will be
reported shortly \cite{jaime}.)  One sees here the stiffening of the equation
of state from inclusion of three-body forces, slightly mitigated by
relativistic effects.  Figure~\ref{akmal-structure}a shows the gravitational
mass vs. central density for families of stars calculated by integrating the
Tolman-Oppenheimer-Volkoff equation for the same equation of state as in
Fig.~\ref{akmal-pnm}, with beta equilibrium of the nucleons included.  The
maximum mass for the nucleonic equation of state, A18+$\delta$v+UIX, is
$\simeq$ 2.2 $M_{\odot}$, marginally consistent with observed neutron star
masses.  By constrast, without three-body forces, the maximum mass is $\sim
1.6 M_{\odot}$, below some observed masses.  The corresponding mass vs. radius
of the families of models is shown in Fig.~\ref{akmal-structure}b; the radii
of these models vary little with mass, and are in the range 10-12 km, except
at the extremes.

    An equation of state based on nucleon interactions alone, while accurately
describing neutron star matter in the neighborhood of $n_0$, has several
fundamental limitations.  One should not expect beyond a few times $n_0$ that
the forces between particles can be described in terms of static few-body
potentials.  Since the characteristic range of the nuclear forces is $\sim
1/2m_\pi$, the parameter measuring the relative importance of three and higher
body forces is of order $n/(2m_\pi)^3 \sim 0.4n/fm^{3}$, so that at densities
well above $n_0$ a well defined expansion in terms of two-, three-, four-,
$\dots$, body forces no longer exists.  The nucleonic equation of state
furthermore does not take into account the rich variety of hadronic ($\Delta$,
hyperonic, mesonic, etc.) and quark degrees of freedom in the nuclear system
which become important with increasing density.  Nor can one continue to
assume at higher densities that the system can even be described in terms of
well-defined ``asymptotic'' laboratory particles.  As one sees in Fig.
\ref{akmal-structure}a, the density in the central cores rises well above
$n_0$; equations of state and neutron star models based on consideration of
nuclear matter alone should not be regarded as definitive.

\section{Quark matter and quark droplets}

    Nuclear matter is expected to turn into a quark-gluon plasma
at sufficiently high baryon density.  Figure~\ref{akmal-structure}a shows
effects of including quark matter cores, naively calculated in the simple MIT
bag model, with bag parameter $B$ =122 and 200 MeV/fm$^3$.  Because of the
well-known technical problems in implementing lattice gauge theory
calculations at non-zero baryon density, we do not to date have a reliable
estimate of the transition density at zero temperature or even a compelling
answer as to whether there is a sharp phase transition or a crossover.
Lattice approaches have included a canonical framework \cite{deforcrand},
which suggests a phase transition at $n\sim 3n_0$ in the hadronic phase
to $\sim 10 n_0$ in the quark phase; and the density of states method
\cite{sfk}, which yields a transition at baryon chemical potential $\mu_b
\sim$ 750MeV, as well as giving a triple point in the phase diagram at finite
temperature.  See also \cite{ejiri}.  Representative field-theoretic
calculations have been carried out in effective NJL theories \cite{ruster}; in
strong coupling qcd \cite{kawamoto}, and in terms of instanton overlap
\cite{zhit}.  Although estimates of the density range of the transition, $\sim
5-10 n_0$, are possibly above the central density found in neutron stars
models based on nuclear equations of state, the question of whether the
dominant degrees of freedom of the matter in the deep cores of neutron stars
are quark-like remains open.  In the absence of information about the equation
of state at very high densities, the issue of whether a distinct family of
quark stars with higher central densities than neutron stars can exist also
remains open.

    If pure neutron matter and quark matter are distinct phases with a first
order transition between them, the transition occurs at nucleonic density
$n_q$ where the energy per baryon of quark matter crosses below that in
neutron matter.  However, the transition in neutron matter with a small
admixture of protons and electrons in beta equilibrium must proceed through a
mixed phase \cite{NG} starting at density below $n_q$.  The mixed phase should
consist of large droplets of quark matter immersed in a sea of hadronic matter
\cite{HPS,PS93}.  Formation of droplets is favored because the presence of s
and d quarks allows reduction of the electron density, and hence electron
Fermi energy, and because it consequently permits an increase in the proton
concentration in the hadronic phase.  The onset of the droplet phase could,
for favorable model parameters of the quark phase, be at a density as low as
$\sim 2n_0$.  A typical droplet is estimated to have a radius of $\sim 5$ fm,
and contain $\sim 100$ u, and $\sim 300$ d as well as s quarks, and thus
having a net negative charge $\sim$ 150, but the results are very model
dependent.

\section{Neutron star masses}

    Observations of neutron star masses constrain the equation of state in the
cores of neutron stars.  The general rule obeyed by families of neutron stars
generated at various central densities from a given equation of state is that
the stiffer the equation of state, the higher is the maximum mass that a
neutron star can have, but the lower is the central mass density, $\rho_c$ at
the maximum mass.  Lower central density means that there is less room for
exotic matter in the interior including $\pi$ and K meson condensates, as well
as quark matter.  Observations of millisecond binary radio pulsars, consisting
of two orbiting neutron stars, have permitted accurate determinations of their
neutron star masses, as well as confirmed the existence of gravitational
radiation; the masses lie in a relatively narrow interval, $\sim 1.35 \pm 0.04
M_{\odot}$ \cite{thorsett}, a mass reminiscent of the Chandrasekhar core mass
of the pre-supernova star.  Were the maximum neutron star mass of order $1.4
M_{\odot}$, the central densities could be sufficiently large to allow
substantial exotica in the interior.

    Even though the range of measured masses of neutron stars in binary
neutron-star systems is tightly constricted, not all neutron stars must have
such small masses.  The constriction reflects the narrow evolutionary track
that allows the two neutron stars to remain bound after their predecessors
undergo supernova explosions \cite{willems}.  A number of determinations of
late of neutron star masses in compact binary x-ray sources call into question
whether the maximum mass is indeed of order $1.4 M_{\odot}$.  The first is
that of the neutron star in the x-ray binary, Vela X-1, with mass deduced to
lie in the range $1.86 \pm 0.33 M_{\odot}$ \cite{barziv,kerkwijk}.  (Also
\cite{quaintrell} which infers 1.75$M_\odot<M<2.44M_\odot$.)  The
uncertainties in the measurement arise from uncertainties in the dynamic
behavior of the atmosphere of the B-supergiant companion star, HD 77581; while
the reported mass, if confirmed, would rule out very soft equations of state,
e.g., those based on kaon condensation, the uncertainties in the determination
do not allow one to make a definitive conclusion.  A second measurement is
that of the neutron star mass in the low mass x-ray binary, Cyg X-2, 1.78$\pm
0.23 M_{\odot}$ \cite{orosz}.  Such higher masses in x-ray binaries would
allow for some exotic matter to be present in neutron stars.

    More recently Nice et al.  \cite{nice} have reported mass determination of
the neutron star in the 3.4ms pulsar PSR J0751+1807 in a close circular (6h)
binary orbit about a helium white dwarf.  The measured neutron star mass is
2.1$\pm 0.2M_\odot$, almost at the limit of compatibility with the nucleonic
equation of state!  The companion mass is 0.191$\pm0.015M_\odot$.  While white
dwarfs do not give rise to the uncertainties present in the Vela X-1
companion, the thermal structure of the white dwarf irradiated by the pulsar
is incompletely understood \cite{nicecrust}.

    Another very promising approach to measuring neutron star masses is
through understanding the origins of the KHz quasiperiodic oscillations
(QPO's) observed in low mass x-ray binary neutron star systems \cite{vdk}.
The power spectra of these sources are characterized by pairs of peaks with a
nearly constant frequency difference.  The QPO's arise from gas accreted from
the companion star working its way through a disk down to orbits very close to
the neutron star surface, where general relativistic effects are crucial.  If,
as is strongly suggested by detailed models \cite{mlp}, the upper QPO
frequency $\nu_2$ is that of orbital motion of the accreted matter in the
innermost circular stable orbit about the neutron star, at radius $R_{\rm
ISCO} = 6MG/c^2 = 3R_{\rm Schwarzschild}$, then one directly infers $M=
c^3/12\sqrt6\,\pi G\nu_2$.  For the QPO system 4U1820-30, with $\nu_2 \simeq$
1070 Hz \cite{zhang}, one would deduce a mass $\simeq 2.0 M_\odot$.

    Finally, \"Ozel, analyzing the neutron star in the low mass x-ray binary
EXO0748-676, a thermonuclear burst source, finds a mass $\simeq 2.1\pm
0.28M_\odot$, and radius $R \simeq 13.8 \pm 1.8$km -- on the outer edge of the
radii predicted by the Akmal et al. equation of state.  These results suggest
that the equation of state may be even stiffer at lower densities, $\simle
4n_0$.

\section{Conclusions}

    Observations of masses close to the maximum, $\simeq 2.2 M_\odot$,
predicted by the nucleonic equation of state begin to challenge our knowledge
of the physics of neutron star interiors.  The existence of high mass neutron
stars immediately indicates that the equation of state must be very stiff,
whether produced by interacting nucleons or other physics.  Central densities
are unlikely to be well above $\sim 7n_0$.  If there is a sharp deconfinement
phase transition below this density, then neutron stars could have quark
matter cores as long as the quark matter is itself very stiff.  Could
$M_{max}$ be larger?  Given the equation of state up to a mass density
$\rho_f$, the maximum possible mass is produced if the equation of state above
$\rho_f$ has sound velocity $c_s = \sqrt{\partial P/\partial\rho}$ equal to
the speed of light \cite{ruffini}.  The $c_s$ predicted by the nucleonic
equation of state \cite{akmal} reaches $c$ at $n \sim 7n_0$; modifying the
physical input, e.g., by including further degrees of freedom such as
hyperons, mesons, or quarks at densities in this neighborhood would lower the
energy per baryon and tend to decrease the stiffness.  Larger $M_{max}$ would
require larger $c_s$ at lower densities; e.g., if one maintains the nuclear
equation of state only up to $\rho_f = 2\rho_0$, then the maximum mass can be
as large as 2.9 $M_\odot$ \cite{vicky}.  Such stiffening would lead to larger
radii, perhaps more consistent with that reported for the neutron star in
EXO0748-676 \cite{ozel}.

    Nonetheless, quarks degrees of freedom -- not accounted for by interacting
nucleons interacting via static potential -- are expected play a role in
neutron stars.  As nucleons begin to overlap, quark degrees of freedom should
become more important, as stressed by Horowitz \cite{horowitz}.  Indeed, once
nucleons overlap considerably the matter should percolate, opening the
possibility of their quark constituents propagating throughout the system
(although near the onset of percolation valence quarks may prefer to remain
bound in triplets, mimicking nucleons, and leaving the matter a color
insulator) \cite{perc,satz}.  Furthermore, the transition from hadronic to
quark matter at low temperature is likely a crossover from BCS-paired
superfluid hadronic matter to superfluid quark matter \cite{colorsup,chiral}.
A firm assessment of the role of quarks in neutron stars must await a better
understanding of mechanisms of quark deconfinement with increasing baryon
density.

\begin{theacknowledgments}

    I would like to dedicate this talk to the memory of my dear friend and
colleague, Vijay Pandharipande, who contributed so much to my understanding of
neutron star interiors.  This work was supported in part by NSF Grant No.
PHY03-55014.

\end{theacknowledgments}


\begin{thebibliography}{99}

    \bibitem{review} G. Baym, and C. Pethick, \emph{ Ann.  Rev.  Nucl.  Sci.}
{\bf 25} (1975) 27, \emph{ Ann.  Rev.  Astron.  Astrophys.} {\bf 17} (1979)
415; C. J. Pethick and D. G. Ravenhall, \emph{ Ann.  Rev.  Nucl.  Part.  Sci.}
{\bf 45}, 429 (1995).

    \bibitem{pion} A.B.  Migdal, \emph{Rev.  Mod. Phys} {\bf 50} (1978) 107;
G.E.  Brown and W. Weise, {\it Phys.  Rept.} {\bf 27} (1976) 1; G. Baym and D.
K. Campbell, in \emph{Mesons in Nuclei}, v. 3, edited by M. Rho and D.
Wilkinson, North-Holland Publ.~Co., Amsterdam, 1979, p. 1031.

    \bibitem{kaon} D. B. Kaplan and A. E. Nelson, \emph{Phys.  Letters} {\bf
B175} (1986) 57; G. E. Brown, K. Kubodera, M. Rho, and V. Thorsson,
\emph{Phys.  Letters} {\bf B291} (1992) 355.

    \bibitem{WFF} R. B. Wiringa, V. Fiks, and A. Fabrocini, \emph {Phys.
Rev. C} {\bf 38} (1988) 1010.

    \bibitem{akmal} A. Akmal, V.R.  Pandharipande, and D.G.  Ravenhall, \emph
{Phys.  Rev. C} {\bf 58}, 1804 (1998).

    \bibitem{pieper} S.C.  Pieper, R.B.  Wiringa, and J. Carlson, \emph{ Phys.
Rev.  C} {\bf 70}, 054325 (2004).

    \bibitem{IL3b} S. C. Pieper, V. R. Pandharipande, R. W. Wiringa and J.
Carlson, \emph{Phys.  Rev.  C}{\bf 64}, 014001 (2001).

    \bibitem{jaime} J. Morales and D.G.  Ravenhall, to be published.

    \bibitem{deforcrand} Ph. de Forcrand and S. Kratochvila, hep-lat/0602024;
and these proceedings.

    \bibitem{sfk} C. Schmidt, Z. Fodor, and S. Katz, hep-lat/0510087,0512032;
C. Schmidt, hep-lat/0610116.

    \bibitem{ejiri} S. Ejiri, F. Karsch, E. Laermann, and C. Schmidt,
\emph{Phys.  Rev.  D}{\bf 73}, 054506 (2006).

    \bibitem{ruster} S.B.  R\"uster, V. Werth, M. Buballa, I. A. Shovkovy, and
D. H. Rischke, nucl-th/0602018.

    \bibitem{kawamoto} N. Kawamoto, K. Miura, A. Ohnishi, and T. Ohnuma,
hep-lat/0512023.

    \bibitem{zhit} A.R.  Zhitnitsky, hep-ph/0601057.

    \bibitem{NG} N. Glendenning, \emph {Phys.  Rev.  D} {\bf 46} (1992) 1274.

    \bibitem{HPS} H. Heiselberg, C. J. Pethick, and E. F. Staubo, \emph{ Phys.
Rev.  Letters} {\bf 70} (1993) 1355.

    \bibitem{PS93} V. R. Pandharipande and E. F. Staubo, in \emph{Int.  Conf.
on Astrophys.}, edited by B. Sinha, World Scientific, Singapore, 1993.

    \bibitem{thorsett} S.E.~Thorsett and D.~Chakrabarty, \emph{Ap.  J.} {\bf
512}, 288 (1999).

    \bibitem{willems} B. Willems and V. Kalogera, \emph{Ap.  J. Letters} {\bf
603}, L101 (2004).

    \bibitem{barziv} O. Barziv, L. Kaper, M. H. van Kerkwijk, J. H. Telting,
and J. Van Paradijs, \emph{Astron. Astrophys.} {\bf 377} 925 (2001).

    \bibitem{kerkwijk} M. H. van Kerkwijk, in \emph{Compact Stars:  Quest for
New States of Dense Matter}, edited by D. K. Hong et al., World Scientific,
Singapore, 2004, p. 116; astro-ph/0403489.

    \bibitem{quaintrell} H. Quaintrell et al., \emph{Astron.  Astrophys.} {\bf
401}, 313 (2003).

    \bibitem{orosz} J.A.  Orosz, and E. Kuulkers, \emph{MNRAS} {\bf 305}, 132
(1999).

    \bibitem{nice} D.J.  Nice, E.M.  Splaver, I.H.  Stairs, O. L\"ohmer, A.
Jessner, M. Kramer and J.M.  Cordes, \emph{Ap.  J.} {\bf 634}, 1242 (2005).

    \bibitem{nicecrust} C. G. Bassa, M. H. van Kerkwijk, S. Kulkarni,
astro-ph/0601205.

    \bibitem{vdk} M. van der Klis, \emph{Ann.  Rev.  Astron.  Astrophys.} {\bf
38}, 717 (2000).

    \bibitem{mlp} M.C. Miller, F.K.  Lamb, and D. Psaltis, {\it Astrophys.
J.} {\bf 508}, 791 (1998); {\it Nucl.  Phys.} {\bf B69}, 123 (1999).

    \bibitem{zhang} W. Zhang, A.P.  Smale,, T.E.  Strohmayer, and J.H.  Swank,
{\it Astrophys.  J.} {\bf 500}, L171 (1998).

    \bibitem{ozel} F. \"Ozel, astro-ph/0605106.

    \bibitem{ruffini} C.E.  Rhoades, Jr. and R. Ruffini, \emph{Phys. Rev.
Letters} {\bf 32}, 324 (1974).

    \bibitem{vicky} V. Kalogera and G. Baym, \emph{Ap.  J. Letters} {\bf 469}
(1996) L61.

    \bibitem{horowitz} C. Horowitz, these proceedings.

    \bibitem{perc} G. Baym, \emph{Physica} {\bf 96A}, l3l (1979).

    \bibitem{satz} H. Satz, \emph{Rept.  Prog.  Phys.} {\bf 63}, 1511 (2000).

    \bibitem{colorsup} K. Rajagopal, \emph{Nucl.  Phys.  A} {\bf 661} 150c,
(1999); M. Alford, K. Rajagopal, and F. Wilczek, \emph{Nucl.  Phys.  B} {\bf
537}, 443 (1999).

    \bibitem{chiral} T. Hatsuda, M. Tachibana, N. Yamamoto, and G. Baym,
\emph{Phys.  Rev.  Letters} {\bf 97}, 122001 (2006).

\end{thebibliography}
\end{document}